\documentclass[12pt]{article}
\usepackage{graphicx}
\usepackage{amsmath}
\usepackage{longtable}

\begin{document}

\makeatletter
\renewcommand*{\@cite}[2]{{#2}}
\renewcommand*{\@biblabel}[1]{#1.\hfill}
\makeatother

\title{3D Interstellar Extinction Map within the Nearest Kiloparsec}
\author{G.~A.~Gontcharov\thanks{E-mail: georgegontcharov@yahoo.com}}

\maketitle

Pulkovo Astronomical Observatory, Russian Academy of Sciences, Pul\-kov\-skoe sh. 65, St. Petersburg, 196140 Russia


The product of the previously constructed 3D maps of stellar reddening (Gontcharov 2010) and
$R_V$ variations (Gontcharov 2012) has allowed us to produce a 3D interstellar extinction map within the
nearest kiloparsec from the Sun with a spatial resolution of 50 pc and an accuracy of $0.2^m$. This map is
compared with the 2D reddening map by Schlegel et al. (1998), the 3D extinction map at high latitudes by
Jones et al. (2011), and the analytical extinction models by Arenou et al. (1992) and Gontcharov (2009). In
all cases, we have found good agreement and show that there are no systematic errors in the new map
everywhere except the direction toward the Galactic center. We have found that the map by Schlegel
et al. (1998) reaches saturation near the Galactic equator at $E_{(B-V)}>0.8^m$, has a zero-point error
and systematic errors gradually increasing with reddening, and among the analytical models those that
take into account the extinction in the Gould Belt are more accurate. Our extinction map shows that
it is determined by reddening variations at low latitudes and $R_V$ variations at high ones. This naturally
explains the contradictory data on the correlation or anticorrelation between reddening and $R_V$ available
in the literature. There is a correlation in a thin layer near the Galactic equator, because both reddening
and $R_V$ here increase toward the Galactic center. There is an anticorrelation outside this layer, because
higher values of $R_V$ correspond to lower reddening at high and middle latitudes. Systematic differences
in sizes and other properties of the dust grains in different parts of the Galaxy manifest themselves in this
way. The largest structures within the nearest kiloparsec, including the Local Bubble, the Gould Belt,
the Great Tunnel, the Scorpius, Perseus, Orion, and other complexes, have manifested themselves in the
constructed map.

\newpage
\section*{INTRODUCTION}

Previously (Gontcharov 2010, below referred to
as G2010), we showed the possibility of constructing
a 3D map of stellar reddening $E_{(J-Ks)}$ based
on infrared (IR) photometry in the J and $Ks$ bands
for 70 million stars from 2MASS catalogue (Skrutskie
et al. 2006) with the most accurate photometry.
We analyzed the distribution of stars on the
$(J-Ks)$ -- $Ks$ diagram. One of the maxima of this
distribution corresponds to F-type dwarfs and subgiants
widespread in the Galaxy with a mean absolute
magnitude $M_{Ks}=2.5^{m}$. The shift of this maximum
toward large $(J-Ks)$ with increasing $Ks$ reflects the
reddening of these stars with increasing heliocentric
distance $R$. As a result, the mean distance and mean
Ks magnitude are proportional for each spatial cell,
to within a small correction: $\overline{R}=10^{(\overline{Ks}+2.5)/5}$. Distributing
the sample of F-type dwarfs and subgiants
among the $Ks$, $l$, and $b$ cells with statistically significant
numbers of stars in each cell, we distribute them
among the 3D spatial cells. The mean reddening of
these stars in such a cell is determined with respect to
the mean color index of nearby stars: $\overline{E_{(J-Ks)}}=\overline{(J-Ks)}-\overline{(J-Ks)_{0}}$.
The first version of the map
from G2010 presents the reddening $E_{(J-Ks)}$ with
an accuracy of $0.03^{m}$ and a spatial resolution of 100 pc
within 1600 pc of the Sun.

Previously (Gontcharov 2012, below referred to
as G2012), we constructed a map of variations in
the extinction law expressed by the extinction coefficient
$R_V$ within 600 pc of the Sun based on multicolor
photometry from the 2MASS and Tycho-2 (H\o g et al. 2000) catalogues.

Given the proportionality $E_{(B-V)}=1.92E_{(J-Ks)}$ (Rieke and Lebofsky 1985), the product of these
maps allows a 3D interstellar extinction map in the
solar neighborhood to be produced using the formula
\begin{equation}
\label{avrvebv}
A_V=R_V\cdot~E_{(B-V)}.
\end{equation}
This paper is devoted to the construction of such a
map and to its analysis and comparison with other
results.

\section*{COMPARISON OF THE 3D AND 2D MAPS}

\subsection*{Refinement and Correction of the G2010 Map}

All of the results published in G2010 are correct.
However, two errors were made in Table 1 from
G2010:
\begin{itemize}
\item
In all readings on the $Y$ scale, the sign should
be reversed, for example, $+550$ and $-550$ pc
should be replaced by $-550$ and $+550$ pc, respectively.
\item
In the part of the table presented in electronic
form, an error in calculating the zero point
was made when $E_{(J-Ks)}$ was recalculated
from spherical coordinates $l$, $b$, $r$ to rectangular
ones $X$, $Y$, $Z$: in particular, for $|Z|>50$ pc,
$E_{(J-Ks)}<0.02$ should be in all cells with
$E_{(J-Ks)}=0.02$. This applies only to the
table; all the corresponding results of the paper
are correct, in particular, a nonzero mean reddening
was specified correctly near the poles.
\end{itemize}

In this paper, to increase the spatial resolution
of the G2010 map, we use moving averaging of the
same original photometric data reduced by the same
method as that in G2010. The reddening $E_{(J-Ks)}$ is calculated for the same spatial $100\times100\times100$
pc cubes as those in G2010, but each cube is
shifted not by 100 pc but by 50 pc in each of the
rectangular Galactic coordinates $X$, $Y$, or $Z$. Thus,
we construct a map of reddening $E_{(J-Ks)}$ with a
spatial resolution of 50 pc. Since it only refines the
G2010 results, below we will retain its designation as
the G2010 map.

\subsection*{The Local-to-Total Galactic Extinction Ratio}

Let us estimate the ratio of the reddening/extinction
in the solar neighborhood (local extinction) to the
total reddening/extinction along the line of sight
within the Galaxy for various $l$ and $b$ (see Table 1). In
accordance with the Besancon model of the Galaxy
(Robin et al. 2003), we take the Galactocentric
distance of the Sun to be 8.5 kpc and the radius of
the equatorial absorbing layer to be within the disk,
i.e., 14 kpc from the Galactic center.

The ``P, 500 pc'' and ``P, 1300 pc'' columns in the
table give the ratio of the extinction within 500 and
1300 pc of the Sun to the total Galactic one when only
an equatorial absorbing layer with the distribution of
matter from Parenago (1954, p. 265) is present:
\begin{equation}
\label{avp}
A\mathrm v=A_{0}\cdot Z_{A}\cdot(1-e^{-r|\sin(b)|/Z_{A})}/|\sin(b)|,
\end{equation}
where $A_{0}$ is the extinction in the equatorial plane
per kpc, and $Z_{A}$ is the characteristic half-thickness
of the absorbing layer. We will take typical estimates:
$A_{0}=1.5^{m}$, $Z_{A}=0.1$ kpc. The table presents
the result only for $l=0$, because the dependence of
extinction on $l$ does not manifest itself for $|b|>5^{\circ}$. An
uncertainty appears at $b=0$. We see from the table
that the extinction in the immediate solar neighborhood
makes a major contribution to the extinction at
middle and high latitudes.

As was shown in G2010 and G2012, the absorbing
matter is distributed in the solar neighborhood
not only along the Galactic equator but also along
the Gould Belt, whose general description was given
by Perryman (2009, pp. 324--328, and references
therein).

Previously (Gontcharov 2009; below referred to as
G2009), we proposed an analytical 3D model that
added up the extinctions in two layers -- along the
Galactic equator and the Gould Belt. The layers
intersect at an angle $\gamma$ and have characteristic half-thicknesses
$Z_{A}$ and $\zeta_{A}$, respectively. The main plane
of the equatorial layer is shifted relative to the Sun by
a distance $\zeta_{0}$; the analogous shift of the layer of the
Gould Belt is $\zeta_{0}$. The axis of intersection between the
layers is turned relative to the $Y$ axis through an angle
$\lambda_{0}$. The longitude $\lambda$ and latitude $\beta$ of a star in the
coordinate system of the Gould Belt are calculated as
\begin{equation}
\label{equ1}
\sin(\beta)=\cos(\gamma)\sin(b)-\sin(\gamma)\cos(b)\cos(l)
\end{equation}
\begin{equation}
\label{equ2}
\tan(\lambda-\lambda_{0})=\cos(b)\sin(l)/(\sin(\gamma)\sin(b)+\cos(\gamma)\cos(b)\cos(l)).
\end{equation}
The extinction $A_V$ is approximated by the sum of two
functions:
\begin{equation}
\label{aaa}
A\mathrm v=A(r,l,b)+A(r,\lambda,\beta),
\end{equation}
each representing a modified formula (2). The extinction
in the equatorial layer is
\begin{equation}
\label{aeq}
A(r,l,b)=(A_{0}+A_{1}\sin(l+A_{2}))Z_{A}(1-e^{-r|\sin(b)|/Z_{A}})/|\sin(b)|,
\end{equation}
where $A_{0}$, $A_{1}$, $A_{2}$ are the free term, the extinction
amplitude, and phase in the sinusoidal dependence
on $l$. The extinction in the Gould Belt is
\begin{equation}
\label{ago}
A(r,\lambda,\beta)=(\Lambda_{0}+\Lambda_{1}\sin(2\lambda+\Lambda_{2}))\zeta_{A}(1-e^{-r|\sin(\beta)|/\zeta_{A}})/|\sin(\beta)|,
\end{equation}
where $\Lambda_{0}$, $\Lambda_{1}$, $\Lambda_{2}$ are the free term, the extinction
amplitude, and phase in the sinusoidal dependence on
$2\lambda$. The assumption that the extinction in the Gould
Belt has two maxima in the dependence on longitude
$\lambda$ was confirmed both in G2009 and in this study.
The extinction maxima in the Gould Belt are observed
near the directions where the distance of the Belt from
the Galactic equator is maximal, i.e., approximately at
the longitudes of the Galactic center and anticenter.

Given the shift of the Sun relative to the absorbing
layers, Eqs. (6) and (7) transform to
\begin{equation}
\label{aeq2}
A(r,l,Z)=(A_{0}+A_{1}\sin(l+A_{2}))r(1-e^{-|Z-Z_{0}|/Z_{A}})Z_{A}/|Z-Z_{0}|,
\end{equation}
\begin{equation}
\label{ago2}
A(r,\lambda,\zeta)=(\Lambda_{0}+\Lambda_{1}\sin(2\lambda+\Lambda_{2}))r(1-e^{-|\zeta-\zeta_{0}|/\zeta_{A}})\zeta_{A}/|\zeta-\zeta_{0}|,
\end{equation}
where $\zeta$ is an analog of the distance $Z$, the shift of
a star in the coordinate system of the Gould Belt
perpendicular to the equatorial plane of the Belt. The
quantities $|Z-Z_{0}|/Z_{A}$ and $|\zeta-\zeta_{0}|/\zeta_{A}$, which are
encountered twice in these formulas, are the characteristics
of the stellar position in the absorbing layers
shifted relative to the Sun.

In G2009, we used individual extinctions for tens
of thousands of stars from three catalogs to determine
the most probable model parameters. The solution
found (12 unknowns) is presented in Table 4, in the
``G2009'' column, and below is compared with the
solutions obtained here.

In accordance with this solution, the two columns
of Table 1 designated as ``G2009, 500 pc'' and
``G2009, 1300 pc'' give the ratio of the extinction
within 500 and 1300 pc of the Sun to the total
Galactic one from the G2009 model, i.e., we adopt the
extinction in both layers (the sum of Eqs. (6) and (7))
within 400 pc and only the extinction in the equatorial
layer (Eq. (6)) farther from the Sun. The results
are not symmetric relative to the Galactic equator
and the Galactic center--anticenter direction due to
the extinction variations in the G2009 model with
longitudes in the planes of the layers. In this model,
the extinction in the immediate solar neighborhood
also makes a major contribution to the extinction at
middle and high latitudes.


\begin{table*}[!h]
\def\baselinestretch{1}\normalsize\footnotesize
\caption[]{Local-to-total Galactic extinction ratio as a function of $l$, $b$,
and the radius of the solar neighborhood under
consideration for two models: Parenago (1954) (designated as ``P'') and G2009
}
\label{pg}
\[
\begin{tabular}{l|rrrrrrrrrr}
\hline
\noalign{\smallskip}
   & \multicolumn{1}{c|}{P, 500 pc} &  \multicolumn{1}{c|}{P, 1300 pc} & \multicolumn{4}{c|}{G2009, 500 pc} & \multicolumn{4}{c|}{G2009, 1300 pc}  \\
\hline
 $b$/$l$ & 0$^{\circ}$ & 0$^{\circ}$ & 0$^{\circ}$ & 90$^{\circ}$ & 180$^{\circ}$ & 270$^{\circ}$ & 0$^{\circ}$ & 90$^{\circ}$ & 180$^{\circ}$ & 270$^{\circ}$ \\
\hline
\noalign{\smallskip}
 70$^{\circ}$ & 0.99 & 1.00 & 1.00 & 1.00 & 1.00 & 1.00 & 1.00 & 1.00 & 1.00 & 1.00 \\
 50$^{\circ}$ & 0.97 & 1.00 & 1.00 & 1.00 & 1.00 & 1.00 & 1.00 & 1.00 & 1.00 & 1.00 \\
 30$^{\circ}$ & 0.92 & 1.00 & 1.00 & 1.00 & 1.00 & 1.00 & 1.00 & 1.00 & 1.00 & 1.00 \\
 25$^{\circ}$ & 0.88 & 1.00 & 1.00 & 0.99 & 0.99 & 0.99 & 1.00 & 1.00 & 1.00 & 1.00 \\
 20$^{\circ}$ & 0.83 & 0.99 & 0.98 & 0.96 & 0.97 & 0.97 & 1.00 & 1.00 & 1.00 & 1.00 \\
 15$^{\circ}$ & 0.73 & 0.97 & 0.95 & 0.90 & 0.93 & 0.92 & 1.00 & 1.00 & 1.00 & 1.00 \\
 10$^{\circ}$ & 0.58 & 0.90 & 0.86 & 0.77 & 0.84 & 0.81 & 1.00 & 0.99 & 0.99 & 1.00 \\
  5$^{\circ}$ & 0.35 & 0.68 & 0.61 & 0.52 & 0.62 & 0.57 & 0.88 & 0.85 & 0.87 & 0.86 \\
  0$^{\circ}$ &      &      & 0.02 & 0.04 & 0.17 & 0.05 & 0.05 & 0.11 & 0.30 & 0.11 \\
 $-5^{\circ}$ & 0.35 & 0.68 & 0.53 & 0.48 & 0.64 & 0.52 & 0.80 & 0.79 & 0.85 & 0.80 \\
$-10^{\circ}$ & 0.58 & 0.90 & 0.75 & 0.71 & 0.86 & 0.74 & 0.94 & 0.94 & 0.97 & 0.94 \\
$-15^{\circ}$ & 0.73 & 0.97 & 0.86 & 0.83 & 0.95 & 0.84 & 0.98 & 0.97 & 0.99 & 0.97 \\
$-20^{\circ}$ & 0.83 & 0.99 & 0.92 & 0.90 & 0.97 & 0.90 & 0.99 & 0.98 & 1.00 & 0.98 \\
$-25^{\circ}$ & 0.88 & 1.00 & 0.94 & 0.93 & 0.98 & 0.93 & 0.99 & 0.99 & 1.00 & 0.99 \\
$-30^{\circ}$ & 0.92 & 1.00 & 0.96 & 0.95 & 0.98 & 0.95 & 0.99 & 0.99 & 1.00 & 0.99 \\
$-50^{\circ}$ & 0.97 & 1.00 & 0.98 & 0.98 & 0.99 & 0.98 & 0.99 & 0.99 & 1.00 & 0.99 \\
$-70^{\circ}$ & 0.99 & 1.00 & 0.99 & 0.99 & 0.99 & 0.99 & 1.00 & 1.00 & 1.00 & 1.00 \\
\hline
\end{tabular}
\]
\end{table*}


In the case of significant changes in the characteristics
of the absorbing layers from Eqs. (6) and (7),
the results in Table 1 will also change. Below, we
recalculate the characteristics of the layers when approximating
our 3D extinction map by the G2009
model. The recalculated characteristics turn out to be
so close to those in G2009 that the data in Table 1 for
$|b|>5^{\circ}$ will change by no more than 0.01 when using
them.

Some of the quantities that are not described by
the models under consideration are known much
more poorly: the extent of the absorbing layers, their
possible warps and thickness variations, and the
Galactocentric distance of the Sun. However, our
calculations showed that the local-to-total Galactic
extinction ratio under consideration also changes
here by less than 0.01 for latitudes $|b|>5^{\circ}$ at any
reasonable values of the radius of the Gould Belt
(from 300 to 600 pc), the radius of the equatorial layer
(from 10 to 23 kpc from the Galactic center), and the
Galactocentric distance of the Sun (from 6 to 9 kpc).

As we see from Table 1, almost all of the absorbing
matter is at a distance of less than 500 pc from the
Sun at high and middle latitudes ($|b|>15^{\circ}$, more
than 70\% of the sky) and closer than 1300 pc for
$10^{\circ}<|b|<15^{\circ}$ (another 10\% of the sky). This allows
the G2010 reddening map to be compared with
the most detailed and justified 2D reddening map by
Schlegel et al. (1998; below referred to as SFD98).

\subsection*{The SFD98 Map}

The SFD98 map presents the IR emission of dust
over the entire sky at a wavelength of 100 microns as a
function of the Galactic longitude and latitude. Data
from the COBE/DIRBE and IRAS/ISSA space
projects served as the original material for mapping.
When the SFD98 map was constructed, a number of
procedures were performed. As a result, the accuracy
of the COBE/DIRBE data (16\%) was combined with
the IRAS/ISSA angular resolution (about 6 arcmin). Given
the dust temperature and the adopted calibrations,
once the zodiacal light and bright point sources have
been eliminated, the IR emission of dust must correspond
to the reddening of stars when their light
passes through the entire Galactic matter on a given
line of sight. Multicolor photometry and spectroscopy
for several hundred elliptical galaxies were used to
calibrate the reddening $E_{(B-V)}$ based on IR emission.
To use the result obtained as an interstellar
extinction map, the authors of SFD98 adopted a constant
extinction coefficient, $R_V=3.1$.

Owing to its high accuracy and angular resolution,
the SFD98 map has been used in many studies (more
than 6000 references!), especially intensively in estimating
the reddening and extinction for extragalactic
objects. Using it to estimate the extinction within the
Galaxy is complicated by the fact that the distances
to the numerous structures reflected on the map are
unknown. Among them, there are regions of extremely
high and low reddening as well as filamentary
structures, including those at high Galactic latitudes.
However, the SFD98 map is \emph{two-dimensional} and,
when compared with the SFD98 map, the \emph{three-dimensional}
map presented in this study will allow
the distances to the most important extinction-related
Galactic structures to be estimated.

In addition, since the construction of the SFD98
map, various authors have discussed its systematic
and random errors. This study allows them to be
estimated.

For zones $\Delta b=5^{\circ}$, Table 2 gives the mean differences
$\overline{\Delta E_{(B-V)}}$ of the reddening from the G2010
map for stars at distances from 1000 to 1600 pc
(below designated as $E_{(B-V)G}$) and the reddening
from the SFD98 map (below designated as $E_{(B-V)SFD98}$) as well as the corresponding standard deviations
$\sigma(E_{(B-V)G}-E_{(B-V)SFD98})$. Such a
wide range of distances for $E_{(B-V)G}$ was chosen to
smooth out its fluctuations due to the small number
of stars. Such a range is admissible, because at a
distance from 1000 to 1600 pc far from the equator,
say, at $|b|<5^{\circ}$, the distance $Z$ from the Galactic
equator will be from 90 to 140 pc and the reddening
and extinction at such heights change little, while the
increase in reddening and extinction near the equator
farther than 1600 pc exceeds considerably that in the
range 1000--1600 pc.

We see from the table that $E_{(B-V)G}$ systematically
exceeds $E_{(B-V)SFD98}$ everywhere except
the four zones nearest to the equator. In addition,
we see that $\sigma(E_{(B-V)G}-E_{(B-V)SFD98})$ at $|b|>15^{\circ}$ nowhere exceeds 0.07$^{m}$.
This corresponds to the
declared high accuracy of the G2010 map ($\sigma(E_{(B-V)G})<0.06^{m}$) and points to a high accuracy of the
SFD98 map ($\sigma(E_{(B-V)SFD98})<0.04^{m}$).

The scatter of differences $\Delta E_{(B-V)}$ increases
sharply near the equator. Here, the mean differences
$\overline{\Delta E_{(B-V)}}$ are negative. This corresponds to the
above calculations of the local-to-total Galactic extinction
ratio (Table 1): as expected, the G2010 map
(showing the reddening within about 1600 pc of the
Sun) near the equator gives considerably lower values
than the SFD98 map (showing the reddening when
light passes through the entire Galaxy).


\begin{table}[!h]
\def\baselinestretch{1}\normalsize\footnotesize
\caption[]{Mean differences $\overline{\Delta E_{(B-V)}}$ and standard deviations
$\sigma(\Delta E_{(B-V)})$ of the reddening from the G2010
map for stars at distances from 1000 to 1600 pc and the
SFD98 map for 5$^{\circ}$-wide $b$ bands
}
\label{bggsfd}
\[
\begin{tabular}{rrr}
\hline
\noalign{\smallskip}
    $b$, deg  & $\Delta E_{(B-V)}$ & $\sigma(\Delta E_{(B-V)})$ \\
\hline
\noalign{\smallskip}
87.5$^{\circ}$  & 0.05  & 0.05 \\
82.5$^{\circ}$  & 0.06  & 0.06 \\
77.5$^{\circ}$  & 0.07  & 0.06 \\
72.5$^{\circ}$  & 0.06  & 0.04 \\
67.5$^{\circ}$  & 0.07  & 0.04 \\
62.5$^{\circ}$  & 0.07  & 0.04 \\
57.5$^{\circ}$  & 0.08  & 0.04 \\
52.5$^{\circ}$  & 0.08  & 0.04 \\
47.5$^{\circ}$  & 0.08  & 0.03 \\
42.5$^{\circ}$  & 0.10  & 0.04 \\
37.5$^{\circ}$  & 0.11  & 0.03 \\
32.5$^{\circ}$  & 0.09  & 0.03 \\
27.5$^{\circ}$  & 0.10  & 0.04 \\
22.5$^{\circ}$  & 0.12  & 0.06 \\
17.5$^{\circ}$  & 0.08  & 0.07 \\
12.5$^{\circ}$  & 0.05  & 0.11 \\
7.5$^{\circ}$   & $-0.05$ & 0.11 \\
2.5$^{\circ}$   & $-0.21$ & 0.16 \\
$-2.5^{\circ}$  & $-0.22$ & 0.16 \\
$-7.5^{\circ}$  & 0.00  & 0.20 \\
$-12.5^{\circ}$ & 0.10  & 0.20 \\
$-17.5^{\circ}$ & 0.09  & 0.07 \\
$-22.5^{\circ}$ & 0.10  & 0.05 \\
$-27.5^{\circ}$ & 0.11  & 0.05 \\
$-32.5^{\circ}$ & 0.09  & 0.07 \\
$-37.5^{\circ}$ & 0.12  & 0.04 \\
$-42.5^{\circ}$ & 0.11  & 0.04 \\
$-47.5^{\circ}$ & 0.09  & 0.04 \\
$-52.5^{\circ}$ & 0.09  & 0.04 \\
$-57.5^{\circ}$ & 0.07  & 0.04 \\
$-62.5^{\circ}$ & 0.07  & 0.05 \\
$-67.5^{\circ}$ & 0.07  & 0.04 \\
$-72.5^{\circ}$ & 0.05  & 0.05 \\
$-77.5^{\circ}$ & 0.07  & 0.04 \\
$-82.5^{\circ}$ & 0.05  & 0.06 \\
$-87.5^{\circ}$ & 0.04  & 0.05 \\
\hline
\end{tabular}
\]
\end{table}


The same difference between the local and Galactic
reddenings at low latitudes is seen as dark regions
in Fig. 1, where the differences $E_{(B-V)G}-E_{(B-V)SFD98}$ are shown as a function of $l$ and $b$. The
black tone corresponds to a difference of $-0.5^{m}$. The
isoline step is 0.1$^{m}$. The vast regions of two predominant
gray tones in the figure correspond to $-0.05^{m}<E_{(B-V)G}-E_{(B-V)SFD98}<0.15^{m}$. We see that
the discrepancy between the maps at middle and high latitudes does not exceed 0.25$^{m}$. The mean difference
is $0.06^{m}$. These differences are large enough to assume the existence of systematic differences between the maps.

Far from the equator, the dark tones in the figure
corresponding to $E_{(B-V)G}<E_{(B-V)SFD98}$ are
seen in the cloud complexes of the Gould Belt: the
mid-latitude extension of the
Aquila rift ($l\approx15^{\circ}$, $b\approx+15^{\circ}$),
Cepheus ($l\approx110^{\circ}$, $b\approx+17^{\circ}$),
Perseus and Orion ($l\approx180^{\circ}$, $b\approx-30^{\circ}$),
Chamaeleon ($l\approx300^{\circ}$, $b\approx-16^{\circ}$),
$\rho~Oph$ ($l\approx354^{\circ}$, $b\approx+18^{\circ}$). In these
regions, the SFD98 study revealed the maximum
dust temperature gradient for middle latitudes and, as
was shown by Arce and Goodman (1999), the resolution
of the dust temperature map accompanying the
SFD98 map is only $1.4^{\circ}$ rather than 6 arcmin, as for the main
map. This should lead to a deviation of the SFD98
map from the true one in both directions at a steep
temperature gradient. Indeed, several regions with
$E_{(B-V)G}>E_{(B-V)SFD98}$ (light spots) are also
noticeable near the mentioned regions and the Lupus
complex ($l\approx340^{\circ}$, $b\approx+13^{\circ}$).

The same effect apparently plays a role in the lightest
regions in Fig. 1, $E_{(B-V)G}>E_{(B-V)SFD98}$. In fact, they form a ringlike region around
the direction toward the Galactic center, but they do
not affect the center itself (the light tones are especially
pronounced at $l\approx0^{\circ}$, $b\approx-10^{\circ}$ -- this region
is discussed below). The dust temperature gradient
is also steep here. It apparently did not allow the
SFD98 map to properly reflect the increase in reddening
toward the Galactic center. At the same time,
the G2010 map is saturated in these sky regions due
to the limitation of the mapping method: the method
is efficient only at $E_{(B-V)}<0.8^{m}$, because at high
reddenings the F-type dwarfs and subgiants used in
the method are mixed with red giants and red dwarfs
on the $(J-Ks)$ -- $Ks$ diagram. Closer to the Galactic
center and at $|b|<5^{\circ}$, the SFD98 map is also saturated:
although the possibility of such saturation was
not pointed out by the authors directly, it is suggested
both by the distribution of values with a jump near
$E_{(B-V)}\approx0.8^{m}$ and by the acknowledgment of the
SFD98 authors that the influence of a large number
of point IR sources is not ruled out near the Galactic
equator. Thus, both maps deviate noticeably from the
true one near the direction toward the Galactic center.

All of these features are also seen in Fig. 2, where
the correlation between $E_{(B-V)G}$ and $E_{(B-V)SFD98}$ is shown for $10^{\circ}\times 10^{\circ}$ sky fields. The
random errors are indicated for all data in the figure:
0.16\% for SFD98 and 0.06$^{m}$ for G2010 (obviously,
the systematic errors somewhere exceed these estimates).

The data for $|b|>15^{\circ}$ outside the clouds of the
Gould Belt are indicated by the filled circles (the main
``swarm'' of points is in the lower left corner of the
figure). These were fitted by a polynomial using the
least-squares method:
\begin{equation}
y=3x^3-3.7x^2+1.8x+0.06,
\label{gsfd}
\end{equation}
where $y$ is $E_{(B-V)G}$, $x$ is $E_{(B-V)SFD98}$.
Thus, there exists a nonlinear relation between
$E_{(B-V)G}$ and $E_{(B-V)SFD98}$: the SFD98 map
underestimates the reddening at its low values, in
particular, near the poles, by $0.06^m$ and overestimates
it at $E_{(B-V)}>0.3^{m}$. The latter effect was detected
by several authors (for references, see Cambresy
et al. 2005). Arce and Goodman (1999) explain
it by the fact that SFD98 used few galaxies with
$E_{(B-V)}>0.15^{m}$ when calibrating $E_{(B-V)}$ from
galaxies and, as a result, the SFD98 calibration at
high reddenings is in error by a factor of 1.3--1.5.
Cambresy et al. (2005) explain this effect by the
properties of dust particles. All estimates from the
literature are in good agreement with the data in
Fig. 2, but a smooth systematic dependence is seen
in this paper for the first time. The underestimation
of the reddening near the poles by the SFD98 map
is also explainable: the zero point of the SFD98
reddening scale is based on the assumption that there
is no reddening near the poles.

The data for nine regions with $|b|>15^{\circ}$ containing
the clouds of the Gould Belt are indicated in the figure
by the large filled diamonds. These were fitted by
the dependence indicated by the dash--dotted curve
at the top of the figure. We see that here, as has been
pointed out above, the SFD98 map underestimates
significantly the reddening, while the G2010 map is
close to saturation. It is important that the systematic
dependence of $E_{(B-V)G}$ on $E_{(B-V)SFD98}$ here is
parallel to dependence (10).

The large open diamonds in Fig. 2 indicate the data
for the region around the Galactic center ($-30^{\circ}<l<+30^{\circ}$, $|b|<15^{\circ}$).
Here, the G2010 map is saturated
almost everywhere (the open diamonds at the
top of the figure), while the SFD98 map is saturated
only near the Galactic center (the open diamonds in
the upper right corner of the figure).

The squares in Fig. 2 indicate the data for $|b|<15^{\circ}$ far from the direction toward the Galactic center.
The saturation of the SFD98 map mentioned above
is noticeable (the squares at the right edge of the
figure). The dotted curve was drawn by the leastsquares
method. The vertical shift of this curve from
curve (10) is explained by the difference between the
local (G2010) and total Galactic (SFD98) reddenings.
The parallelism of all three curves in the figure
suggests that the systematic error of the SFD98 map
is the same over the entire sky.

Thus, the G2010 map is apparently more accurate
than the SFD98 map in systematic terms and has a
more accurate zero point (better estimates the reddening
toward the poles), but the SFD98 map is more
accurate near the Galactic center.

We will return to the data in Fig. 2 below when
comparing our results with the 3D extinction map by
Jones et al. (2011).

\section*{EXTINCTION AS THE PRODUCT OF REDDENING AND $R_V$}

According to Eq. (1), the product of the G2010
stellar reddening map and the map of $R_V$ variations
with allowance made for the relation $E_{(B-V)}=1.92E_{(J-Ks)}$ (Rieke and Lebofsky 1985) gives a
map of interstellar extinction $A_V$.

The G2010 reddening map gives reliable data
within about 1600 pc of the Sun at $|Z|<600$ pc
(farther from the Galactic equator, the reddening
depends almost entirely on the zero point determined
with a large relative error -- the reddening toward the
Galactic poles). The G2012 map of $R_V$ variations
gives reliable data only within no more than 600 pc
of the Sun. However, the G2012 data suggest that
the influence of local Galactic structures, the Gould
Belt and the Local Bubble, with which the large
$R_V$ variations near the Sun are apparently associated
ceases at a greater distance. Consequently, at a
greater distance, we can assume $R_V\approx3.1$ far from
the Galactic equator and a monotonic increase in this
quantity along the $X$ coordinate toward the Galactic
center in a thin layer (less than 150 pc in thickness)
near the Galactic equator (G2012):
\begin{equation}
R\mathrm v=2.96+0.00025X.
\label{rvx}
\end{equation}
At the expense of some reduction in accuracy and
reliability, we thus extend the G2012 map of $R_V$ variations
to the $2\times2$ kpc region centered on the Sun:
we adopt it as is within 600 pc of the Sun, dependence
(11) in the layer $|Z|<100$ pc, and $R_V=3.1$
at the remaining locations. Accordingly, we use the
G2010 reddening map and construct the extinction
map in the $2\times2$ kpc region centered on the Sun (remembering
that the accuracy of the results decreases
with increasing heliocentric distance).

Figure 3 shows the contour maps of reddening
$E_{(B-V)}$ (left column), coefficient $R_V$ (central column),
and extinction $A_V$ (right column) as a function
of the $X$ and $Y$ coordinates in the following layers:
(a) $+150<Z<+250$ pc,
(b) $+50<Z<+150$ pc,
(c) $-50<Z<+50$ pc,
(d) $-150<Z<-50$ pc,
(e) $-250<Z<-150$ pc.
The black
tone corresponds to $E_{(B-V)}=0.04^{m}$, $R_V\le2$, $A_V=0.1^{m}$.
The isoline step is $\Delta E_{(B-V)}=0.07^{m}$, $\Delta A_V=0.2^{m}$.
$R_V\ge4$ is marked by the white tone.
The white lines of the coordinate grid are plotted
with a 500-pc step. The Sun is at the centers of the
plots. The Galactic center is on the right. Similar
maps are shown in Figs. 4 and 5, respectively, as
a function of the $X$ and $Z$ coordinates in the layers
(a) $+200<Y<+300$ pc,
(b) $+100<Y<+200$ pc,
(c) $0<Y<+100$ pc,
(d) $-100<Y<0$ pc,
(e) $-200<Y<-100$ pc,
(f) $-300<Y<-200$ pc,
and as a function of the $Y$ and $Z$ coordinates in the layers
(a) $+200<X<+300$ pc,
(b) $+100<X<+200$ pc,
(c) $0<X<+100$ pc,
(d) $-100<X<0$ pc,
(e) $-200<X<-100$ pc,
(f) $-300<X<-200$ pc.

The main thing that is seen on the presented
extinction map is that at low and middle latitudes
($|b|<45^{\circ}$) it is determined mainly by the reddening
map and not by the $R_V$ variations, no matter what
extreme values this coefficient takes on; in contrast,
the extinction at high latitudes is determined mainly
by $R_V$. It is here that $R_V$ reaches its maximum values,
exhibits large variations, and is determined with a
lower accuracy. The role and uncertainty of $R_V$ at
high latitudes are so great that today we even cannot
assert with confidence that $R_V$ does not reach still
higher values still farther from the Galactic plane, in
unstudied regions, due to an increase in the fraction
of coarse dust.

In any case, it is now clear that the low extinction
and reddening at high latitudes in no way contradicts
the high values of $R_V$ there. In addition, the
numerous contradictory data on the correlation or
anticorrelation between reddening and $R_V$ available
in the literature now find an explanation. There is
a correlation in a thin layer ($|Z|<100$ pc) near the
Galactic equator, because both the reddening and $R_V$
increase here toward the Galactic center. Outside this
layer, there is an anticorrelation: at high and middle
latitudes, higher $R_V$ correspond to lower reddening.
Obviously, all of this is determined by systematic
differences in sizes and other properties of the dust
grains in various regions of the Galaxy.

It is also important to estimate the contribution of
the uncertainties in the reddening and $R_V$ to the total
uncertainty in the extinction. Table 3 gives typical
values of $E_{(B-V)}$, $R_V$, and $A_V$ as well as absolute
and relative errors in these quantities toward the
Galactic pole ($|b|\approx90^{\circ}$), at a middle Galactic latitude
($|b|\approx15^{\circ}$), and near the Galactic equator ($|b|\approx0^{\circ}$)
at a heliocentric distance of several hundred pc obtained
in G2010, G2012, and this paper. The relative
error in $A_V$ is equal to the quadratic sum of the relative
errors in $E_{(B-V)}$ and $R_V$. Therefore, we see from
the table that the error in $A_V$ is everywhere determined
by the error in $E_{(B-V)}$, while, according to the
most pessimistic estimates, the relative uncertainty in
the variations of the extinction law, no matter what
its nature is, does not exceed 10\%. Therefore, this
uncertainty does not affect the conclusions reached in
G2010, G2012, and this paper and does not influence
the accuracy of the constructed 3D extinction map at
all latitudes.

In future, however, a refinement of the extinction
law much more laborious than the construction of
a 3D reddening map will come to the fore, because
it will require reproducing the entire wavelength dependence
of extinction. The latter is possible only
by using homogeneous multiband photometry (spectrophotometry
is better) for millions of stars with an
accuracy of at least 0.01$^{m}$ in the ultraviolet, visible,
and IR ranges in many spatial cells of the Galaxy.


\begin{table}
\caption[]{Typical values of $E_{(B-V)}$, $R_V$, and $A_V$ as well as
absolute and relative errors in these quantities at various
latitudes at a heliocentric distance of several hundred pc
obtained in G2010, G2012, and this paper
}
\label{erro}
\[
\begin{tabular}{lccc}
\hline
\noalign{\smallskip}
           Parameter & $|b|\approx90^{\circ}$ & $|b|\approx15^{\circ}$ & $|b|\approx0^{\circ}$ \\
\hline
\noalign{\smallskip}
$E_{(B-V)}$                         & 0.06 & 0.20 & 0.70 \\
$\sigma(E_{(B-V)})$                 & 0.06 & 0.06 & 0.06 \\
$\sigma(E_{(B-V)})/E_{(B-V)}$          & 1.00 & 0.30 & 0.09 \\
\hline
$R_V$                               & 4.20 & 3.10 & 2.80 \\
$\sigma(R_V)$             & 0.20 & 0.20 & 0.20 \\
$\sigma(R_V)/R_V$  & 0.05 & 0.06 & 0.07 \\
\hline
Av                               & 0.25 & 0.62 & 1.96 \\
$\sigma(A_V)$             & 0.25 & 0.19 & 0.22 \\
$\sigma(A_V)/A_V$  & 1.00 & 0.31 & 0.11 \\
\hline
\end{tabular}
\]
\end{table}


Since the main features of the G2010 reddening
map discussed in the corresponding paper are also
retained in the constructed extinction map, let us list
them only briefly.

First of all, the Great Tunnel, a region of low
reddening and extinction stretching in the figure
from the top to the lower left corner, is seen in
Fig. 3. The Great Tunnel was probably first mentioned
by Welsh (1991) as an extension of the
Local Bubble, a region of low gas density within
about 100 pc of the Sun, toward Canis Major ($l\approx230^{\circ}$). Gontcharov (2004) and Gontcharov and
Vityazev (2005) showed that the concentration of
high-luminosity stars, which, in fact, form the Gould
Belt along the Tunnel edges, is enhanced along the
Great Tunnel. Here, the large Scorpius, Perseus,
Orion, and other cloud complexes are located along
the Tunnel edges. All of this emphasizes the role of
the Gould Belt as a dust container. In Fig. 4, the
Gould Belt oriented in this plane at an angle of about
17$^{\circ}$ to the Galactic plane is seen in the orientation of
the largest absorbing structures (the light spots in the
figure).

The region of high reddening and extinction seen
as the white spot in Figs. 1, 3, and 4 ($l\approx0^{\circ}$, $b\approx-10^{\circ}$, $X\approx800$ pc,
$Y\approx0$ pc, $Z\approx-250$ pc, Sagittarius)
has not been identified with any known structures.
This may be an artifact that has emerged from
the limitations of the method in a region with a steep
reddening gradient.

The construction of a 3D extinction map in this
paper offers great prospects for investigating individual
cloud complexes, especially the distances to them
and the relationship between reddening and $R_V$ in
each complex. However, this extensive analysis is
beyond the scope of this publication.

It is also impossible to compare the newly constructed
map with the available numerous results
referring to the extinction in specific regions of space
in one publication. This work will be continued.
Let us consider only one of the most interesting 3D
extinction maps. It refers to the same region of space
as the map obtained here but was constructed by a
completely different method.

\section*{COMPARISON WITH THE 3D EXTINCTION MAP BY JONES et al.}

Jones et al. (2011; below referred to as JWF11)
constructed a 3D extinction map for high latitudes
within 2 kpc of the Sun using the spectra of more
than 9000 M-type dwarfs from the Sloan Digital Sky
Survey (SDSS DR7) (Abazajian et al. 2009). The
extinction was determined for each star by fitting its
spectrum in the range $570<\lambda<920$ nm by a
$\lambda$-dependent extinction curve. Unfortunately, the
data show only the northern polar cap (approximately
$b>50^{\circ}$), several narrow bands in the southern hemisphere,
and isolated small zones of the sky. The data
for 9362 stars from the JWF11 map were compared
with the map constructed here in spherical segments
with sizes $10^{\circ}\times10^{\circ}\times100$ pc in $l$, $b$, and $r$, respectively,
containing at least ten JWF11 stars in each
segment. This number is required to smooth out the
$A_V$ variations from star to star (although the declared
JWF11 accuracy of determining the individual $A_V$
was, on average, 0.04$^{m}$). Only the segments of the
northern polar cap contain a sufficient number of stars
from JWF11.

Our map is in good agreement with that from
JWF11: for the spherical layers $+50<r<+150$, $+150<r<+250$ è $+250<r<+350$ pc, the mean $A_V$ differences between our
study and the JWF11 map are $-0.02^{m}$, $-0.01^{m}$ è $0.02^{m}$, while the standard deviations of the differences
are $0.11^{m}$, $0.07^{m}$ è $0.10^{m}$, respectively. These agree
with the accuracy estimates in this paper and JWF11.
In other regions of space, the data are insufficient for
comparison.

It is important that both maps at high latitudes
give an equally high mean extinction ($\overline{A_V}\approx0.2^{m}$)
compared to the SFD98 map ($\overline{A_V}\approx0.1^{m}$). This confirms
the previously made assumption about the zeropoint
error for the SFD98 map.

The dots in Fig. 6 represent the correlation of
the JWF11 data for individual stars (with $E_{(B-V)}=A_V/3.1$) designated as $E_{(B-V)_JWF11}$ with the
$E_{(B-V)SFD98}$ data, while the solid curve indicates
dependence (10) presented above in Fig. 2 that fits the
correlation between $E_{(B-V)G}$ and $E_{(B-V)SFD98}$
for $10^{\circ}\times10^{\circ}$ and $|b|>15^{\circ}$ sky fields outside the
clouds of the Gould Belt. The dashed curve indicates
relation $E_{(B-V)_JWF11}=E_{(B-V)SFD98}$. We
reached the following conclusions:
\begin{itemize}
\item 
The saturation of the SFD98 map mentioned
above actually exists: the stars with $E_{(B-V)SFD98}>0.8^{m}$ are rare.
\item 
The dots in the lower left corner of the figure,
on the whole, agree with the solid curve,
indicating a nonzero reddening at high latitudes
($\overline{E_{(B-V)}}=0.06^{m}$), as distinct from
the SFD98 map.
\item 
At $E_{(B-V)SFD98}>0.2^{m}$, almost all of the
points lie below the solid curve, reflecting a deviation
of the local reddening for comparatively
close JWF11 stars from the total Galactic one
derived from SFD98.
\item 
At $E_{(B-V)SFD98}<0.2^{m}$, there are many
stars with $E_{(B-V)_JWF11}\gg E_{(B-V)SFD98}$,
which most likely suggests that the errors of
these maps were disregarded at high latitudes.
\item 
On the whole, the cloud of points is parallel
to the solid curve, confirming that the systematic
difference between the G2010 and SFD98
maps indicated by the curve is caused by the
systematic error of the latter.
\end{itemize}

\section*{COMPARISON WITH 3D ANALYTICAL MODELS}

Before the appearance of the G2009 extinction
model, for many years the most widely used analytical
3D interstellar extinction model within the
nearest kiloparsec had been the model by Arenou
et al. (1992), which fits the mean extinction $A_V$ for
199 sky regions by parabolas as a function of the
distance. The shortcoming of this model is the absence
of any physical justification of the extinction
variations.

When comparing our map with the model by Arenou
et al., the mean difference turned out to be zero,
while the standard deviation of the differences was
0.22$^{m}$ for spatial cells closer than 500 pc and 0.19$^{m}$ for
cells closer than 400 pc.

Let us consider the correspondence between our
map and the G2009 model. For this purpose, we
will solve the system of equations (5), one equation
for each spatial cell. The observed extinction $A_V$ is
on the left-hand sides, while the function of $r$, $l$, and
$b$ with the following 12 unknowns are on the righthand
$\gamma$, $\lambda_{0}$, $Z_{A}$, $\zeta_{A}$, $Z_{0}$, $\zeta_{0}$,
$A_{0}$, $A_{1}$, $A_{2}$, $\Lambda_{0}$, $\Lambda_{1}$, $\Lambda_{2}$. They are chosen so as to minimize the sum of
the squares of the residuals of the left- and right-hand
sides of Eqs. (5).

The calculations were performed for spatial cells
within 500 pc of the Sun at $|Z|<300$ pc. Table 4
presents 12 unknowns and the standard deviation of
the map from the model designated as $\sigma$. The ``E''
column provides data only for the equatorial layer with
the characteristics giving the smallest deviation from
the map; the ``E+G'' column provides data for both
layers with the characteristics giving the smallest deviation
from the map. For comparison, the ``G2009''
column presents the result obtained in G2009, while
the $\sigma$ row gives two standard deviations -- when fitting
only by the equatorial layer and by both layers.
The accuracies of determining the unknowns
were estimated from the range of variations in the
unknowns as the standard deviation increased significantly,
by 0.004$^{m}$.

We see that the derived map is better described by
the model with extinction in two layers rather than
one, only equatorial layer.

Among the characteristics found, $A1$ è $A2$ differ
noticeably from those in G2009. In contrast,
the difference between $A_{0}$ and $\Lambda_{0}$ was explained in
G2009, where they are also significantly different in
the solutions using different data: because of the low
inclination of the Gould Belt to the Galactic equator
and, accordingly, the small region of longitudes where
the absorbing layers separate, only the total constant
extinction is determined with confidence. This is also
seen in the new solutions: the sum $A_{0}+\Lambda_{0}$ is almost
the same in three columns of Table. 4 -- 2.1$^{m}$, 2.2$^{m}$ è 2.35$^{m}$.

Thus, the constructed map agrees well with all of
the models considered, but it agrees best with the
G2009 model that allows for the extinction in the
Gould Belt.


\begin{table}
\caption[]{Solutions of the system of equations (5)}
\label{sol}
\[
\begin{tabular}{lccc}
\hline
\noalign{\smallskip}
        Parameter   & G2009 & E & E+G \\
\hline
\noalign{\smallskip}
$\gamma$, deg         & $17$      &              & $19\pm1$     \\
$\lambda_{0}$, deg    & $-10$     &              & $-15\pm5$     \\
$Z_{A}$, pc                & $70$      & $70\pm5$     & $68\pm6$       \\
$Z_{0}$, pc                & $10$      & $-6\pm2$     & $-9\pm2$       \\
$A_{0}, ^{m}/$kpc          & $1.2$     & $2.2\pm0.1$  & $1.9\pm0.1$   \\
$A_{1}, ^{m}/$kpc          & $0.6$     & $0.3\pm0.05$ & $0.3\pm0.05$   \\
$A_{2}$, deg          & $35$      & $54\pm7$     & $55\pm5$      \\
$\zeta_{A}$, pc            & $50$      &              & $40\pm10$     \\
$\zeta_{0}$, pc            & $0$       &              & $5\pm3$       \\
$\Lambda_{0}, ^{m}/$kpc    & $1.1$     &              & $0.45\pm0.07$   \\
$\Lambda_{1}, ^{m}/$kpc    & $0.9$     &              & $1.1\pm0.1$  \\
$\Lambda_{2}$, deg    & $135$     &              & $130\pm10$     \\
$\sigma, ^{m}$             & 0.25/0.18 & 0.20         & 0.17           \\
\hline
\end{tabular}
\]
\end{table}


\section*{CONCLUSIONS}

This is the fourth paper in our series of studies
of the interstellar extinction in the Galaxy. As our
previous studies (G2009, G2010, G2012), it shows
that the stellar reddening and extinction can be analyzed
using accurate multicolor broadband photometry
from present-day surveys of millions of stars and
that the Gould Belt plays a great role as a region
containing absorbing matter in clouds oriented approximately
radially with respect to the Sun.

The product of the stellar reddening map (G2010)
by the map of $R_V$ variations (G2012) allows an extinction
map within the nearest kiloparsec from the
Sun to be constructed with a spatial resolution of
50 pc and an accuracy $\sigma(A_V)=0.2^{m}$. At present,
there are high-resolution and high-accuracy 2D extinction
maps (e.g., SFD98), high-resolution and
high-accuracy 3D maps \emph{for part of the sky} (e.g.,
JWF11), and fairly accurate all-sky 3D maps \emph{farther
than 2 kpc from the Sun}. The map presented here is
currently one of the few 3D maps for the entire nearest
kiloparsec.

Both theoretical calculations and data analysis
point to a great role of the absorbing matter within the
nearest kiloparsec: it determines the extinction in the
bulk of the sky and is particularly important for estimating
the extinction toward extragalactic objects.

The limited volume of the paper forces us to compare
the constructed reddening and extinction maps
primarily with the most popular results: the SFD98
reddening map and the analytical extinction models
by Arenou et al. (1992) and G2009. In all cases,
good agreement was found, but we showed that the
SFD98 map has saturation near the Galactic equator,
systematic errors in regions with a high reddening,
and a zero-point error, while the analytical extinction
model benefits if the extinction in the Gould Belt is
explicitly taken into account.

In the constructed extinction map, we found no
systematic errors nowhere except the region around
the direction toward the Galactic center.

Our extinction map showed that it is determined
mainly by reddening variations at low and middle
latitudes ($|b|<45^{\circ}$) and by $R_V$ variations at high latitudes.
Since it is here that $R_V$ reaches its maximum
values, exhibits large variations, and is determined
with a lower accuracy, its refinement is needed. In this
paper, we explain for the first time the contradictory
data on the correlation or anticorrelation between
reddening and $R_V$ available in the literature. There
is a correlation in a thin layer ($|Z|<100$ pc) near the
Galactic equator, because both the reddening and $R_V$
increase here toward the Galactic center. There is
an anticorrelation outside this layer: higher values of
$R_V$ correspond to lower reddening at high and middle
latitudes. Obviously, all of this is determined by
systematic differences in sizes and other properties of
the dust grains in different regions of the Galaxy.

Since the largest structures within the nearest
kiloparsec, including the Local Bubble, the Gould
Belt, the Great Tunnel, the Scorpius, Perseus, Orion,
and other complexes, have manifested themselves
in the constructed map, its comprehensive analysis
and refinement offer great prospects for investigating
these structures, especially the distances to them and
the local dust properties.

\section*{ACKNOWLEDGMENTS}

In this study, we used data from the Hipparcos and 2MASS (Two Micron AllSky Survey) projects as
well as the SIMBAD database and other resources of
the Strasbourg Data Center (France), http://cds.ustrasbg.fr/. The study was supported by the ``Origin
and Evolution of Stars and Galaxies'' Program of the Presidium of the Russian Academy of Sciences.

\newpage

\begin{figure}
\includegraphics{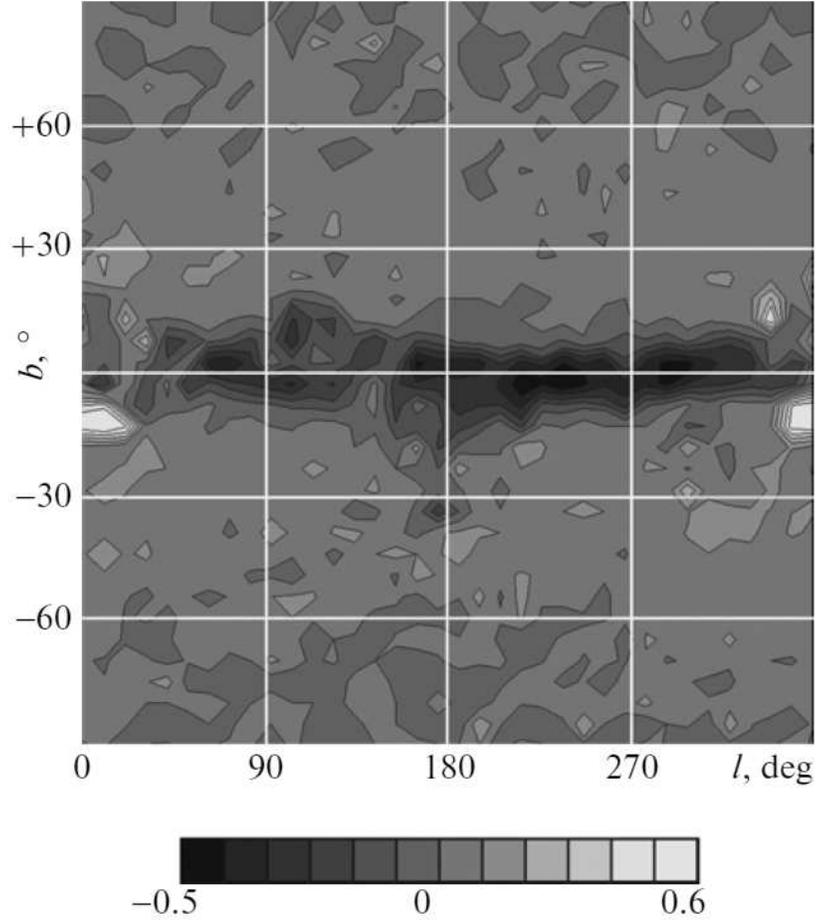}
\caption{Differences between the reddening $E_{(B-V)G}$
from the G2010 map for stars with $1000<r<1600$ pc and $E_{(B-V)SFD98}$ from the SFD98 map as
a function of $l$ and $b$. The black tone corresponds to a
difference of $-0.5^{m}$. The isoline step is 0.1$^{m}$.
}
\label{ggsfd}
\end{figure}

\begin{figure}
\includegraphics{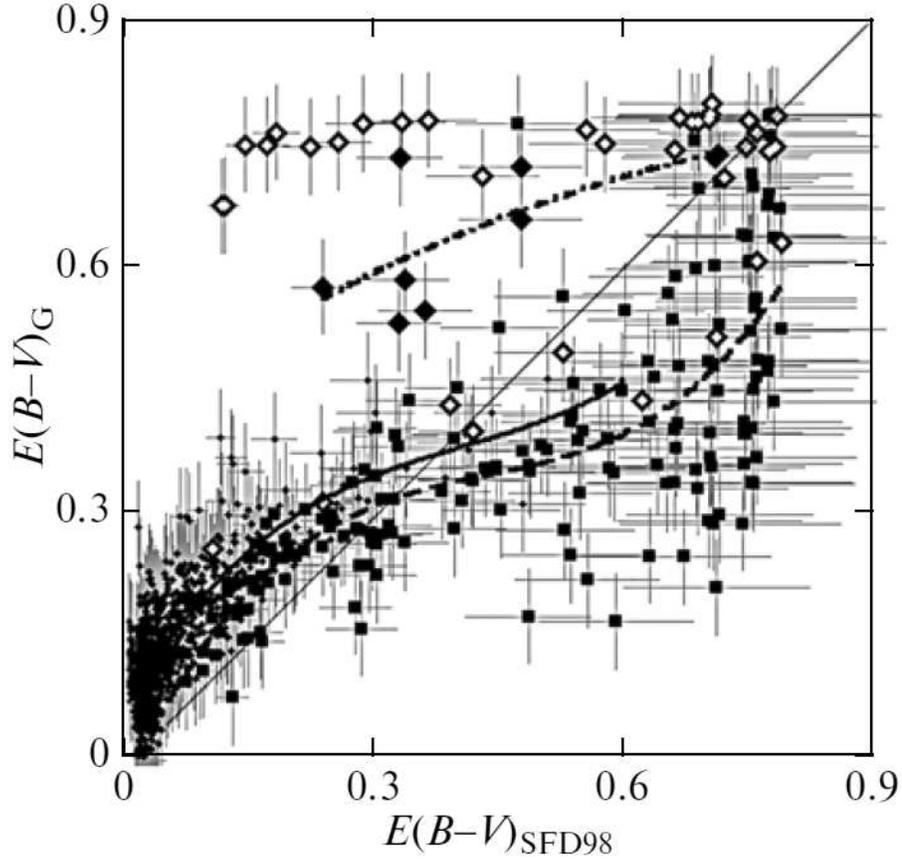}
\caption{Correlation between $E_{(B-V)G}$ and $E_{(B-V)SFD98}$ for $10^{\circ}\times10^{\circ}$ sky fields.
The filled circles indicate
the data for $|b|>15^{\circ}$ outside the clouds of the Gould Belt
fitted by the solid curve. The large filled diamonds indicate
the data for nine regions with $|b|>15^{\circ}$ containing
the clouds of the Gould Belt fitted by the dash-dotted
curve. The large open diamonds indicate the data for the
region around the Galactic center ($-30^{\circ}<l<+30^{\circ}$, $|b|<15^{\circ}$).
The squares indicate the data for $|b|<15^{\circ}$ far
from the direction toward the Galactic center fitted by the
dotted curve.
}
\label{ebvebv}
\end{figure}

\begin{figure}
\includegraphics{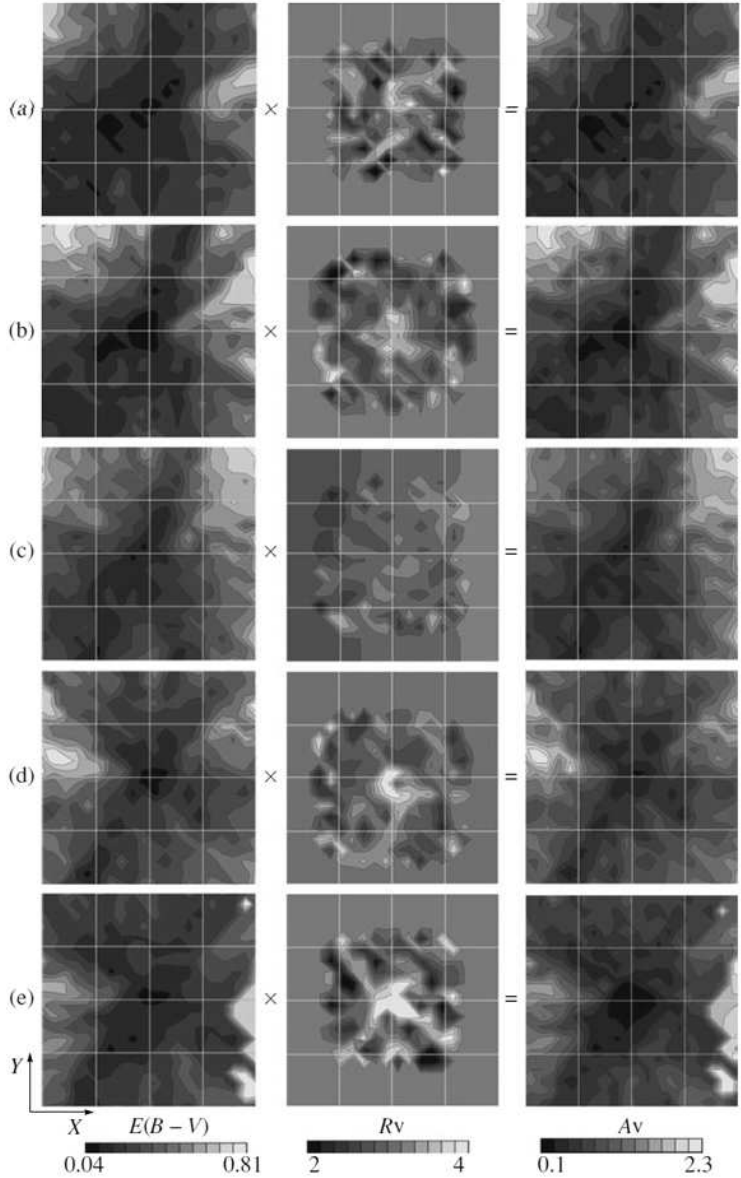}
\caption{Contour maps of reddening $E_{(B-V)}$ (left column), coefficient $R_V$ (central column), and extinction $A_V$
(right column)
as a function of the $X$ and $Y$ coordinates in the following layers:
(a) $+150<Z<+250$ pc,
(b) $+50<Z<+150$ pc,
(c) $-50<Z<+50$ pc,
(d) $-150<Z<-50$ pc,
(e) $-250<Z<-150$ pc. The black tone corresponds to
$E_{(B-V)}=0.04^m$, $R_V\le2$, $A_V=0.1^m$. The isoline step is $\Delta E_{(B-V)}=0.07^{m}$,
$\Delta A_V=0.2^m$. $R_V\ge4$ is marked by the white tone.
The white lines of the coordinate grid are plotted with a 500-pc step.
The Sun is at the centers of the plots. The Galactic center is on the right.
}
\label{xy}
\end{figure}

\begin{figure}
\includegraphics{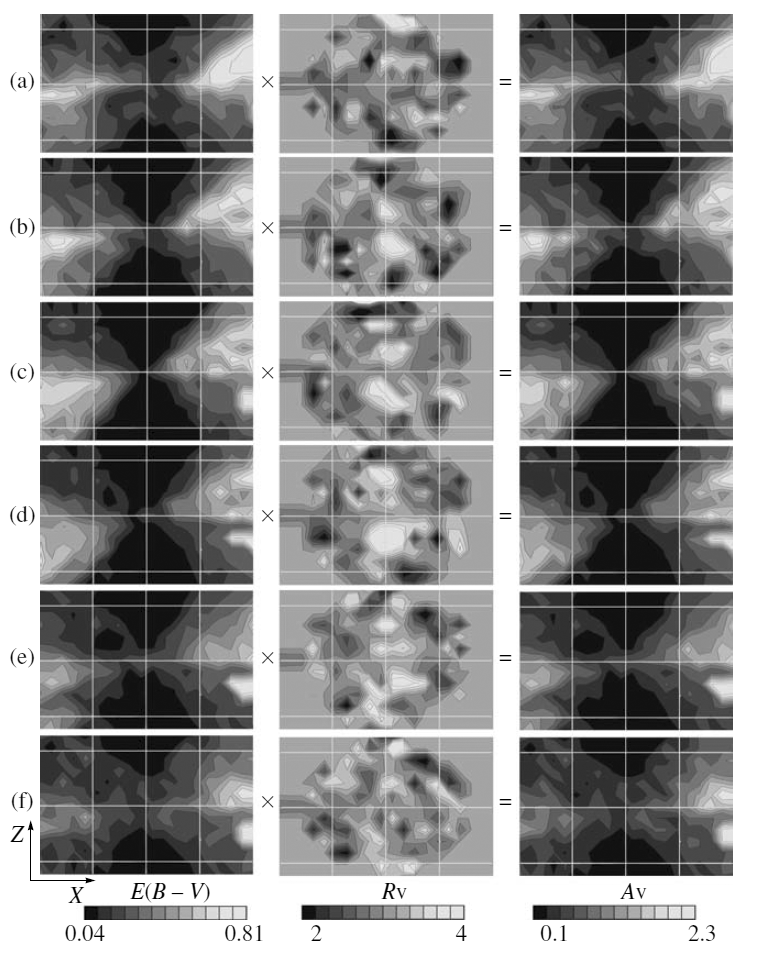}
\caption{Same as Fig. 3, but the maps as a function of the $X$ and $Z$ coordinates in the following layers:
(a) $+200<Y<+300$ pc,
(b) $+100<Y<+200$ pc,
(c) $0<Y<+100$ pc,
(d) $-100<Y<0$ pc,
(e) $-200<Y<-100$ pc,
(f) $-300<Y<-200$ pc.
}
\label{xz}
\end{figure}

\begin{figure}
\includegraphics{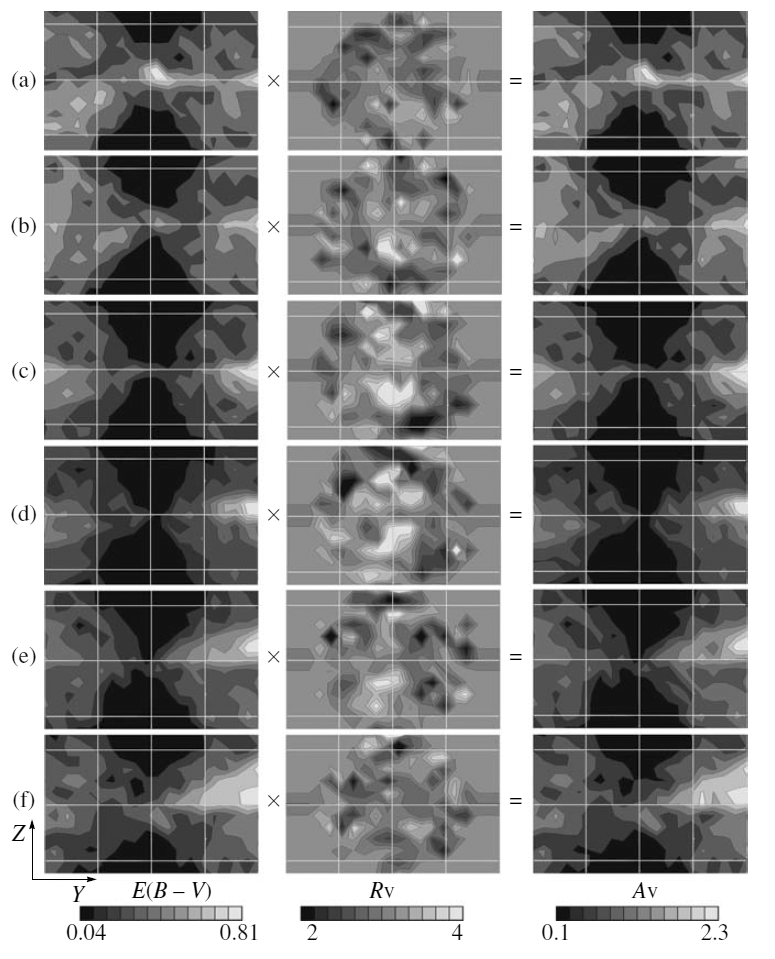}
\caption{Same as Fig. 3, but the maps as a function of the $Y$ and $Z$ coordinates in the following layers:
(a) $+200<X<+300$ pc,
(b) $+100<X<+200$ pc,
(c) $0<X<+100$ pc,
(d) $-100<X<0$ pc,
(e) $-200<X<-100$ pc,
(f) $-300<X<-200$ pc..
}
\label{yz}
\end{figure}

\begin{figure}
\includegraphics{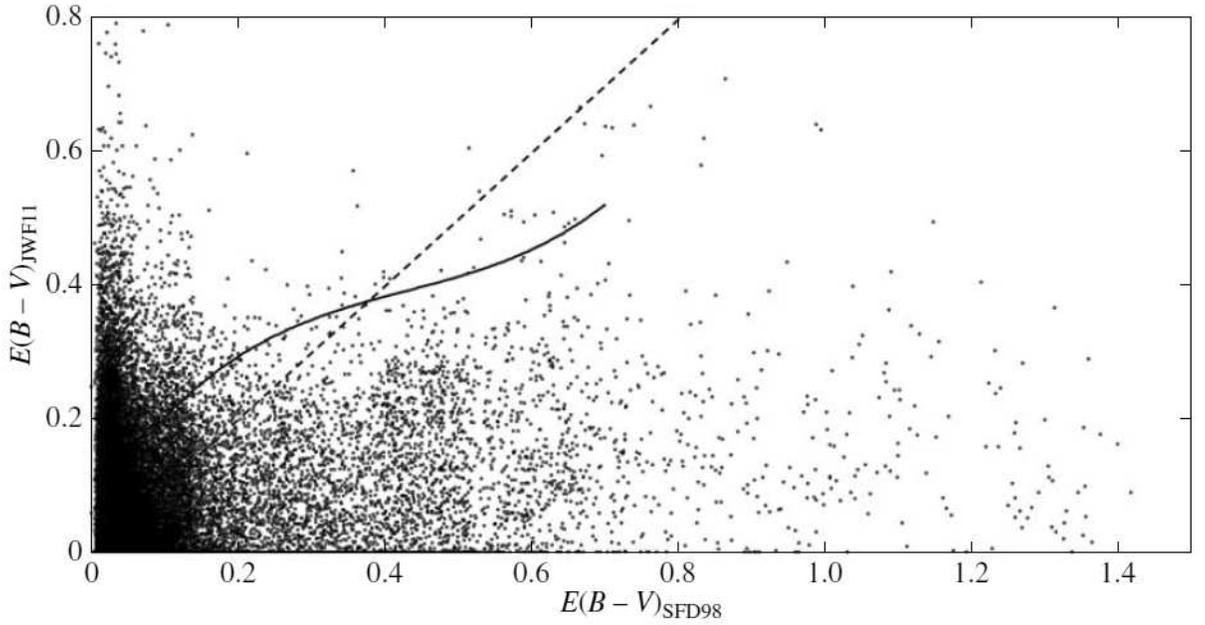}
\caption{The correlation between $E_{(B-V)G}$ and $E_{(B-V)SFD98}$ for $10^{\circ}\times10^{\circ}$,
$|b|>15^{\circ}$ sky fields outside the clouds of the
Could Belt from Fig. 2 is fitted by the solid curve (10). The correlation between $E_{(B-V)}$
for individual stars from the map
by Jones et al. (2011) (with $E_{(B-V)}=A_V/3.1$) and $E_{(B-V)SFD98}$ is indicated by the open squares. The dashed curve
indicates relation 1/1.
}
\label{jon}
\end{figure}

\end{document}